# An improved beam waist formula for ultrashort, tightly-focused linearly, radially, and azimuthally polarized laser pulses in free space

**Liang Jie Wong[1,*], Franz X. Kärtner[2,3], and Steven G. Johnson[1]**

[1] *Department of Mathematics, Massachusetts Institute of Technology, 77 Massachusetts Avenue, Cambridge, MA, 02139, USA*
[2] *Department of Electrical Engineering and Computer Science and Research Laboratory of Electronics, Massachusetts Institute of Technology, 77 Massachusetts Avenue, Cambridge, MA, 02139, USA*
[3] *Center for Free-Electron Laser Science - DESY, and Department of Physics, University of Hamburg, Notkestraße 85, D-22607 Hamburg, Germany*
*\*Corresponding author: ljwong@mit.edu*

We derive an asymptotically accurate formula for the beam waist of ultrashort, tightly-focused fundamental linearly-polarized, radially-polarized, and azimuthally-polarized modes in free space. We compute the exact beam waist via numerical cubature to ascertain the accuracy with which our formula approximates the exact beam waist over a broad range of parameters of practical interest. Based on this, we describe a method of choosing parameters in the model given the beam waist and pulse duration of a laser pulse.

In this letter, we provide a missing piece of an increasingly important exact wave solution for ultrashort, tightly-focused laser pulses, by supplying an asymptotically-accurate analytical expression for the key parameter of the beam waist (in contrast to a previous expression that was valid only for continuous-wave lasers [1]). Ultrashort, tightly-focused laser pulses are employed in a wide range of fields including particle acceleration [2,3], particle trapping [4], X-ray generation [5], and high-resolution microscopy [6]. Exact field solutions are important even in the nominally paraxial, many-cycle regime, as shown by recent studies [7] that reveal severe discrepancies between the off-axis electrodynamics predicted using a paraxial beam model and those predicted using an exact beam model.

Among the various models used to describe non-paraxial electromagnetic pulses [8-11], the model in [10] is especially attractive because it is a closed-form, singularity-free, exact solution to Maxwell's equations. The models in [8,11] are defined in terms of integrals or infinite series. The model in [9], while also a closed-form solution, contains substantial DC components when both Rayleigh range $z_0$ and pulse duration $T$ are small (with $z_0$ and $T$ defined as in [9]).

The beam waist and pulse duration of the model in [10] are controlled by two parameters, typically denoted $a$ and $s$. For paraxial, many-cycle pulses, $a$ is proportional to the square of the beam waist radius and $s$ to the pulse duration. Generally, however, the beam waist radius and pulse duration are each functions of both $a$ and $s$. A method to determine $a$ and $s$ with reasonable accuracy for a given beam waist and pulse duration would lend the model more readily to the description of actual laser pulses. While [10] supplies an asymptotically exact relation between $s$ and the pulse duration, the beam waist formula it cites becomes severely inaccurate at short pulse durations, regardless of the confocal parameter $a$. We derive a single beam waist formula for the lowest order linearly-polarized (LP$_{01}$), radially-polarized (TM$_{01}$), and azimuthally-polarized (TE$_{01}$) modes,

$$w_{\text{pulse}} = \sqrt{\frac{2a}{k_0}} \sqrt{\frac{s}{s+1+\varsigma_0^2/2}}, \qquad (1)$$

which is exact in various limits, and which we show to be empirically highly accurate for all practical values of $a$ and $s$. Specifically, it is exact for the conditions $s \ll 1, s \ll k_0 a$ or $k_0 a \gg 1, \forall s$. In Eq. (1), $\varsigma_0 \equiv 2^{1/2}$ for the LP$_{01}$ mode and $\varsigma_0 \equiv 1$ for the TM$_{01}$ or TE$_{01}$ mode. $k_0 = 2\pi/\lambda_0$, with $\lambda_0$ being the peak wavelength of the pulse.

We first give an overview of the model in [10]. Next, we outline the derivation of Eq. (1) and verify its accuracy by comparing its results with the exact waists computed via numerical cubature over a broad range of parameters of practical interest. Finally, we numerically ascertain that the pulse duration is practically independent of $a$ in the parameter range of interest. This suggests that to determine $a$ and $s$ for a given beam waist and pulse duration, one should first determine $s$ from the pulse duration and then use Eq. (1) to determine $a$.

The model in [10] is based on the complex-source-point method, first introduced by Deschamps to model a beam of electromagnetic radiation using the fields of a point source at an imaginary distance along the propagation axis [12]. Couture and Bélanger showed [13] that the paraxial Gaussian beam becomes the complex-source-point wave when every order of non-paraxial correction is made according to the prescription of Lax et al. [14], making the complex-source-point wave a natural generalization of the Gaussian beam commonly encountered in paraxial optics. One may superpose two counter-propagating solutions to remove the singularities [10,15] inherent in the original model [16-18]. One may also employ a Poisson spectrum to ensure an absence of DC components in the pulsed solution [8,19]. Combining these strategies, [10] uses a spectrum proportional to

$$\tilde{\Psi} = \frac{\sin(kR')}{R'\exp(ka)}\left[\omega^s \exp\left(-\frac{s\omega}{\omega_0}\right)\theta(\omega)\right], \quad (2)$$

where the pre-factor on the right-hand-side is associated with the beam divergence, and the square-bracketed expression (the Poisson spectrum) is associated with the pulse envelope. $R' \equiv (r^2 + (z+ja)^2)^{1/2}$, $j \equiv (-1)^{1/2}$, $k = \omega/c = 2\pi/\lambda$, $\lambda$ is the vacuum wavelength, $a$ is a parameter that affects degree of focusing, $s$ is a parameter related to pulse duration, $\omega_0 = k_0 c = 2\pi c/\lambda_0$ is the peak angular frequency of the Poisson spectrum and $\theta(\cdot)$ is the Heaviside step function. $r$ and $z$ are respectively the radial and longitudinal coordinates of the cylindrical coordinate system. Eq. (2) exactly solves the Helmholtz equation at any $\omega$. Although Eq. (2) does not suffer from singularities, it contains backward-propagating components which a large $a$ is required to suppress. The inverse Fourier transform of Eq. (2) is

$$\Psi = \frac{1}{R'}\left(f_+^{-s-1} - f_-^{-s-1}\right), \quad (3)$$

where $f_\pm = 1 - j/s(\omega_0 t \pm k_0 R' + jk_0 a)$, and we have dropped constant factors. Eq. (3) is then used in the components of the Hertz potentials (which must also satisfy the vector wave equation) to obtain the electric and magnetic fields $\vec{E}$ and $\vec{H}$ in free space. Here, we focus on the lowest-order radially-polarized (TM$_{01}$) and lowest-order linearly-polarized (LP$_{01}$) modes [10]. By the property of electromagnetic duality, the beam waist of the lowest-order azimuthally-polarized (TE$_{01}$) mode is identical to that of the TM$_{01}$ mode.

The continuous-wave (CW) counterpart of the model in [10] is given in [1], where the authors define a beam waist $w_{CW}$ according to the geometry of the oblate spheroidal beam:

$$w_{CW} = \frac{\sqrt{2}}{k_0}\sqrt{\sqrt{1+(k_0 a)^2}-1}. \quad (4)$$

Although it is tempting to use this beam waist formula for the pulsed solution in [10], Eq. (4) in fact approximates the actual waist well only when both pulse duration and waist are large (in which case $w_{CW} \approx w_{pulse} \approx (2a/k_0)^{1/2}$). Counter-intuitively – since $a$ occurs only in the beam factor, and $s$ the pulse factor, of Eq. (2) – the beam waist becomes a strong function of both $a$ and $s$ when $s$ is small. This was noted in [10] but a formula was not provided. The definition we use for the exact beam waist $w_{exact}$, illustrated in Fig. 1, is

$$w_{exact} = \varsigma_0 \sqrt{\frac{\iint r^2 S_z' \, dxdy}{\iint S_z' \, dxdy}}, \quad (5)$$

where the $\sqrt{\cdots}$ factor is the second irradiance moment of the pulse at the focal plane and pulse peak. The double integrals are over the entire focal plane and $S_z'$ is the z-directed Poynting vector component $S_z \equiv \vec{E} \times \vec{H} \cdot \hat{z}$ evaluated at the focal plane, pulse peak and carrier amplitude of the respective mode. $\varsigma_0$ is included so that in the paraxial, CW limit we get fields $\sim \exp(-r^2/w_{exact}^2)$. For the TM$_{01}$ mode,

$$E_r = \text{Re}\left\{A_- \frac{jra}{R'}\right\}, \quad \eta_0 H_\varphi = \text{Re}\left\{-B\frac{r}{R'}\right\}, \quad (6)$$

in the focal plane, where $A_\pm \equiv \partial^2\Psi/\partial^2 R' \pm 1/R' \partial\Psi/\partial R'$ and $B \equiv \partial^2\Psi/\partial(ct)\partial R'$. Substituting Eq. (3) in Eq. (6),

$$\iint r^d S_z' \, dxdy =$$
$$C\int r^{d+1} dr \, \text{Re}\left\{\frac{rja}{R'^2}\left[\frac{3}{k_0 R'^3}g_1 - \frac{3j(s+1)}{sR'^2}g_{2+}\right.\right.$$
$$\left.\left. - \frac{(s+1)(s+2)k_0}{s^2 R'}g_{3-}\right]\right\}. \quad (7)$$
$$\text{Re}\left\{\frac{r}{R'}\left[\frac{j(s+1)}{sR'^2}g_{2-}\right.\right.$$
$$\left.\left. + \frac{(s+1)(s+2)k_0}{s^2 R'}g_{3+}\right]\right\}$$

where $g_1 \equiv f_+^{-s-1} - f_-^{-s-1}$, $g_{2\pm} \equiv f_+^{-s-2} \pm f_-^{-s-2}$, $g_{3\pm} \equiv f_+^{-s-3} \pm f_-^{-s-3}$, $d \in \{0,2\}$, and $C$ is some constant factor. Under the condition $s \ll 1$, we can replace all instances of $s+n, n \in \mathbb{Z}^+$ with $n$. After some tedious algebra, we find that

$$\iint r^d S_z' \, dxdy = C' \int d(R'^2) \frac{(R'^2 + a^2)^{1+d/2}}{(R'^2 + b^2)^6} + O(s), \quad (8)$$

where $b \equiv a + s/k_0$ and $C'$ is some constant factor. Note that $R'^2 = r^2 - a^2 \in \mathfrak{R}$ in the focal plane. Substituting Eq. (8) in Eq. (5), and using the condition $s \ll k_0 a$,

$$k_0 w_{exact} = k_0 \sqrt{\frac{2}{3}(b^2 - a^2)} + O(s^{3/2})$$
$$= \sqrt{\frac{4k_0 as}{3}}\left[1 + O\left(s + \frac{s}{k_0 a}\right)\right], \quad (9)$$
$$= k_0 w_{pulse} + O\left(s^{3/2} + \frac{s^{3/2}}{k_0 a}\right)$$

which agrees with Eq. (1) when $s \ll 1$, $s \ll k_0 a$.

For the case $k_0 a \gg 1$, we use the expansion $R' \approx ja(1 - r^2/2a^2)$ since power transport is significant only where $r \ll a$. From Eq. (7), we have

$$\iint r^d S_z' \, dxdy =$$
$$C'' \int_0^\Delta d\rho \left(2a^2\rho\right)^{d/2} \rho (1+h\rho)^{-2s-6} \left[1 + O\left(\frac{1}{k_0 a}\right)\right], \quad (10)$$

where $h \equiv k_0 a/s$, $\rho \equiv r^2/2a^2$, $\Delta \ll 1$ is a value many times larger than $1/k_0 a$, and $C''$ is some constant factor. Applying Eq. (10) to Eqs. (5) gives

$$k_0 w_{\text{exact}} = \sqrt{2k_0 a} \sqrt{\frac{s}{s+3/2}} + O\left(\frac{1}{\sqrt{k_0 a}}\right)$$
$$= k_0 w_{\text{pulse}} + O\left(\frac{1}{\sqrt{k_0 a}}\right), \quad (11)$$

which reduces to Eq. (1) for the $TM_{01}$ case when $k_0 a \gg 1$. Since the $TE_{01}$ mode is obtained via the changes $\vec{E} \to \vec{H}, \vec{H} \to -\vec{E}, \mu_0 \to \varepsilon_0, \varepsilon_0 \to \mu_0$, in the description of the $TM_{01}$ mode, Eq. (7) also applies for the $TE_{01}$ mode, making the beam waists of the $TM_{01}$ and $TE_{01}$ modes identical. The transverse fields of $LP_{01}$ mode in the focal plane are

$$E_x = \text{Re}\left\{ A_- \frac{x^2}{R'^2} - A_+ + B \frac{ja}{R'} \right\}$$
$$E_y = \eta_0 H_x = \text{Re}\left\{ A_- \frac{xy}{R'^2} \right\} \quad . \quad (12)$$
$$\eta_0 H_y = \text{Re}\left\{ A_- \frac{y^2}{R'^2} - A_+ + B \frac{ja}{R'} \right\}$$

Substituting Eq. (12) in Eq. (5) we find ourselves with integrals in which the integrand is once again azimuthally symmetric, making it convenient to perform the integrations in polar coordinates, as in Eq. (7). One can verify Eq. (1) for the $LP_{01}$ mode by considering the individual cases $s \ll 1, s \ll k_0 a$, and $k_0 a \gg 1$ using the procedure undertaken for the $TM_{01}$ mode in each of the respective cases.

A natural question to ask at this point is whether the pulse duration is also a function of both $a$ and $s$. To answer this question, we define pulse duration $\tau$ as the normalized second moment of time in the laser's focal plane:

$$\tau = \sqrt{\frac{\iiint t^2 S_z \, dxdydt}{\iiint S_z \, dxdydt}}, \quad (13)$$

where the triple integrals are over the entire focal plane and temporal axis, and plot $\tau$ (computed via numerical cubature) as a function of $a$ and $s$ for both the $LP_{01}$ and $TM_{01}$ modes in Fig. 2, where the corresponding plots for $w_0$ are also given. In our simulations, we exclude the region $k_0 a < 5$ since a reasonably large value of $k_0 a$ is required to suppress the backward-propagating components inherent in Eqs. (2) and (3) for the model to be an accurate description of a single forward-propagating pulse.

In Figs. 2(a) and (c) we see that the pulse duration is approximately independent of $a$ in our regime of interest.

Figs. 2(b) and (d) show that the waist is approximately independent of $s$ at large values of $a$, but begins to be strongly affected by $a$ in the few-cycle and sub-cycle pulse regimes ($s = 1$ corresponding roughly to a single-cycle pulse). Since Eq. (4) is not a function of $s$, it is clearly unsuitable as a model of beam waist in this regime. To model a pulsed beam of given beam waist $w_{\text{exact}}$ and pulse duration $\tau$, one should first determine $s$ from $\tau$ [10] and then use $s$ and $w_{\text{exact}}$ to determine $a$ via Eq. (1). If even more accurate approximations to the waist and pulse duration are desired, the values obtained through this procedure may be used as a starting point for manual adjustment or numerical iteration.

In Fig. 3, we examine the accuracies of Eq. (4) and Eq. (1) quantitatively. The relative error of Eq. (4) exceeds 350% at low pulse durations in Fig. 3(b), making Eq. (4) a very poor estimate for short pulses in general. It is interesting, however, that Eq. (4) predicts the waist very accurately along the line $s \approx 2^{-1/2} \varsigma_0 k_0 a$, but the error increases rapidly elsewhere. Note that the relative error in Eq. (1) becomes significant only at low $a$ (and large $s$), where the use of Eqs. (2) and (3) to model a single propagating pulse is unreliable in the first place due to the significant presence of backward-propagating components.

In conclusion, we have derived a beam waist formula, Eq. (1), that is asymptotically accurate at small $s$ and/or large $a$. We have verified that the pulse durations of pulsed $LP_{01}$, $TM_{01}$, and $TE_{01}$ free space modes are approximately independent of $a$ in our regime of interest ($k_0 a > 5$). The beam waist, however, is significantly affected by both $s$ and $a$, and the beam waist formula Eq. (4) derived for a continuous-wave beam is thus a poor estimate of the actual beam waist at small pulse durations, even at large $a$. To model a pulsed $LP_{01}$, $TM_{01}$ or $TE_{01}$ mode in free space given waist radius $w_0$ and pulse duration $\tau$, one should first determine $s$ from $\tau$ and then use $s$ and $w_{\text{exact}}$ to determine $a$ via Eq. (1).

LJW acknowledges support from the Agency for Science, Technology and Research (A*STAR), Singapore.

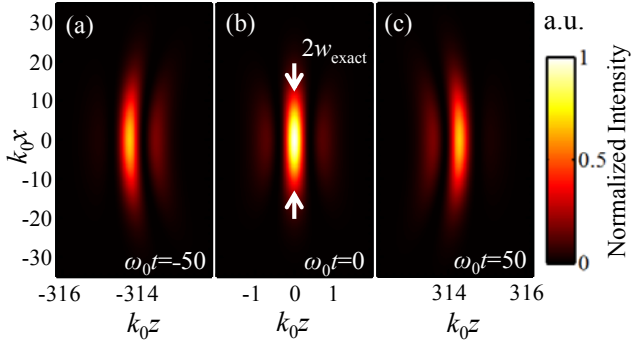

Fig. 1: Color maps of the Poynting vector's z-component for a single-cycle $LP_{01}$ pulse of parameters $k_0 a = 500$ and $s = 1$ at various propagation times. The arrows in (b) indicate the waist diameter (corresponding radius defined in Eq. (5)). Note that the second irradiance moment is evaluated on the focal plane at the pulse peak and carrier peak. The waist radius for this particular case is about $2.9\,\lambda$.

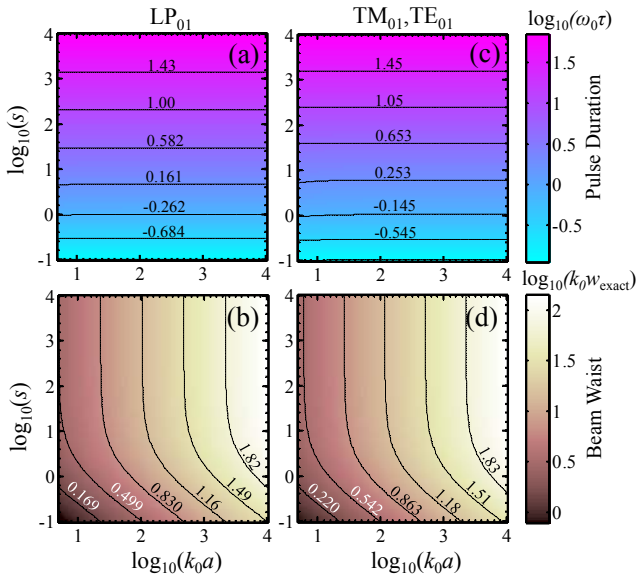

Fig. 2: Color maps of numerically-obtained (a): exact pulse duration $\tau$ and (b) exact beam waist $w_{exact}$ of the $LP_{01}$ mode as a function of $a$ and $s$. (c) and (d) contain the same information as (a) and (b) respectively, except for the $TM_{01}$ (or $TE_{01}$) mode. The dotted lines are contour lines, which show relative invariance of pulse duration with $a$, especially at large $a$, in (a) and (b).

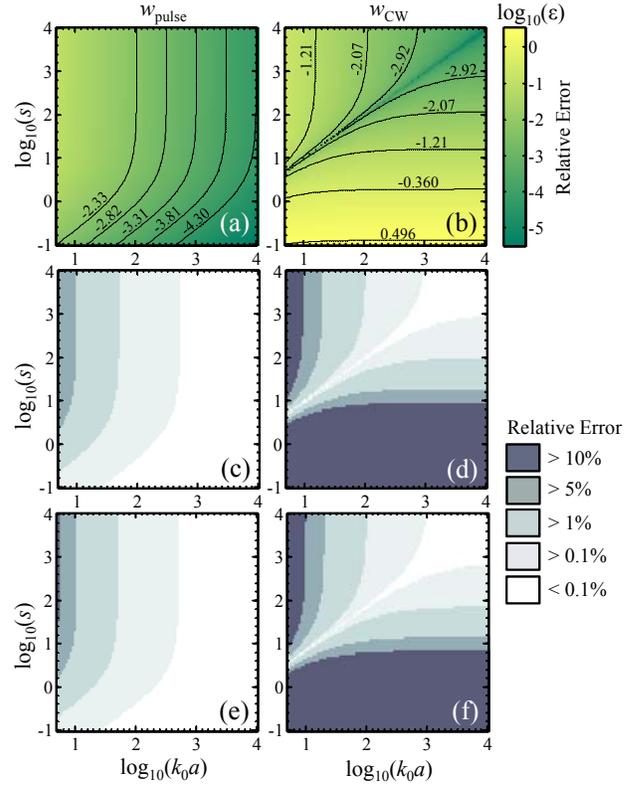

Fig. 3: Relative error $\varepsilon$ of the beam waist formulas in (a) Eq. (1) and (b) Eq. (4) for the $LP_{01}$ mode. (c) and (d) present the information in (a) and (b) respectively in a different form. (e) and (f) correspond to (c) and (d) for the $TM_{01}$ (or $TE_{01}$) mode. The color maps corresponding to (a) and (b) for the $TM_{01}$ (or $TE_{01}$) mode are visually very similar and thus omitted.